\documentclass{INTERSPEECH2023}
\usepackage{float}
\interspeechcameraready

\title{Robust Audio Anti-Spoofing with Fusion-Reconstruction Learning \\on Multi-Order Spectrograms}
\name{Penghui Wen$^1$, Kun Hu$^{1,*}$\thanks{*Corresponding Author. This study was partially supported by Australian Research Council (ARC) grant \#DP210102674. Our code is available at \url{https://github.com/ph-w2000/S2pecNet}.}, Wenxi Yue$^1$, Sen Zhang$^1$, Wanlei Zhou$^2$, Zhiyong Wang$^1$}
\address{
  $^1$School of Computer Science, The University of Sydney, NSW, Australia\\
  $^2$Faculty of Data Science, City University of Macau, Macao SAR, China }
\email{pwen5103@uni.sydney.edu.au, kun.hu@sydney.edu.au, wenxi.yue@sydney.edu.au,  szha2609@uni.sydney.edu.au, wlzhou@cityu.mo,  zhiyong.wang@sydney.edu.au}

\begin{document}

\maketitle
 
\begin{abstract}
Robust audio anti-spoofing has been increasingly challenging due to the recent advancements on deepfake techniques. While spectrograms have demonstrated their capability for anti-spoofing, complementary information presented in multi-order spectral patterns have not been well explored, which limits their effectiveness for varying spoofing attacks. Therefore, we propose a novel deep learning method with a spectral fusion-reconstruction strategy, namely $\text{S}^2\text{pecNet}$, to utilise multi-order spectral patterns for robust audio anti-spoofing representations. 
Specifically, spectral patterns up to second-order are fused in a coarse-to-fine manner and two branches are designed for the fine-level fusion from the spectral and temporal contexts. A reconstruction from the fused representation to the input spectrograms further reduces the potential fused information loss. Our method achieved the state-of-the-art performance with an EER of 0.77\% on a widely used dataset - ASVspoof2019 LA Challenge. 
\end{abstract}

\section{Introduction}

Audio based automatic speaker verification (ASV) has a wide range of applications due to the biometric authentication property of voice \cite{zhang2022norm}, including multi-speaker speech recognition \cite{xu2020speaker}, speech authentication-based telephone banking and voice-based forensics \cite{reynolds1995speaker}. To improve the robustness of ASV systems against increasingly complex deepfake techniques, various methods have been devised~\cite{chen2020generalization,wang2021comparative, sanchez2015toward} to identify audio spoofing.

The identification of audio spoofing is generally treated as a binary classification task which classifies an audio recording as genuine or spoofed. 
Early studies primarily focused on devising hand-crafted features to anti-spoofing, such as Cochlear Filter Cepstral Coefficient Instantaneous Frequency (CFCCIF) \cite{patel2015combining}, Linear Frequency Cepstral Coefficients (LFCC) \cite{sahidullah2015comparison} 
and Constant-Q Cepstral Coefficients (CQCC) \cite{todisco2016new}. 
Recently, various deep learning methods have been proposed for audio deepfake recognition. For example, a convolutional neural network (CNN) was first adopted with 2D audio spectrograms \cite{zhang2017investigation}, then the deep residual network (ResNet) methods (e.g., \cite{zhang2021one}) were proposed to formulate the anti-spoofing problem as one-class feature learning \cite{khan2010survey} to improve the generalisability. 
Based on the effectiveness of sub-band spectrogram features in anti-spoofing, a dual-band fusion algorithm was proposed \cite{zhang122021effect}. To characterise temporal relations in audio signals, a recurrent neural network (RNN) based method was proposed in \cite{gomez2019light}. 
More recently, a spectro-temporal graph attention network method AASIST ~\cite{jung2022aasist} was proposed to formulate spoofing patterns with $1^\text{st}$-order spectrograms (i.e., raw spectrograms). 

\begin{figure}[h]
  \centering
  \includegraphics[width=\linewidth]{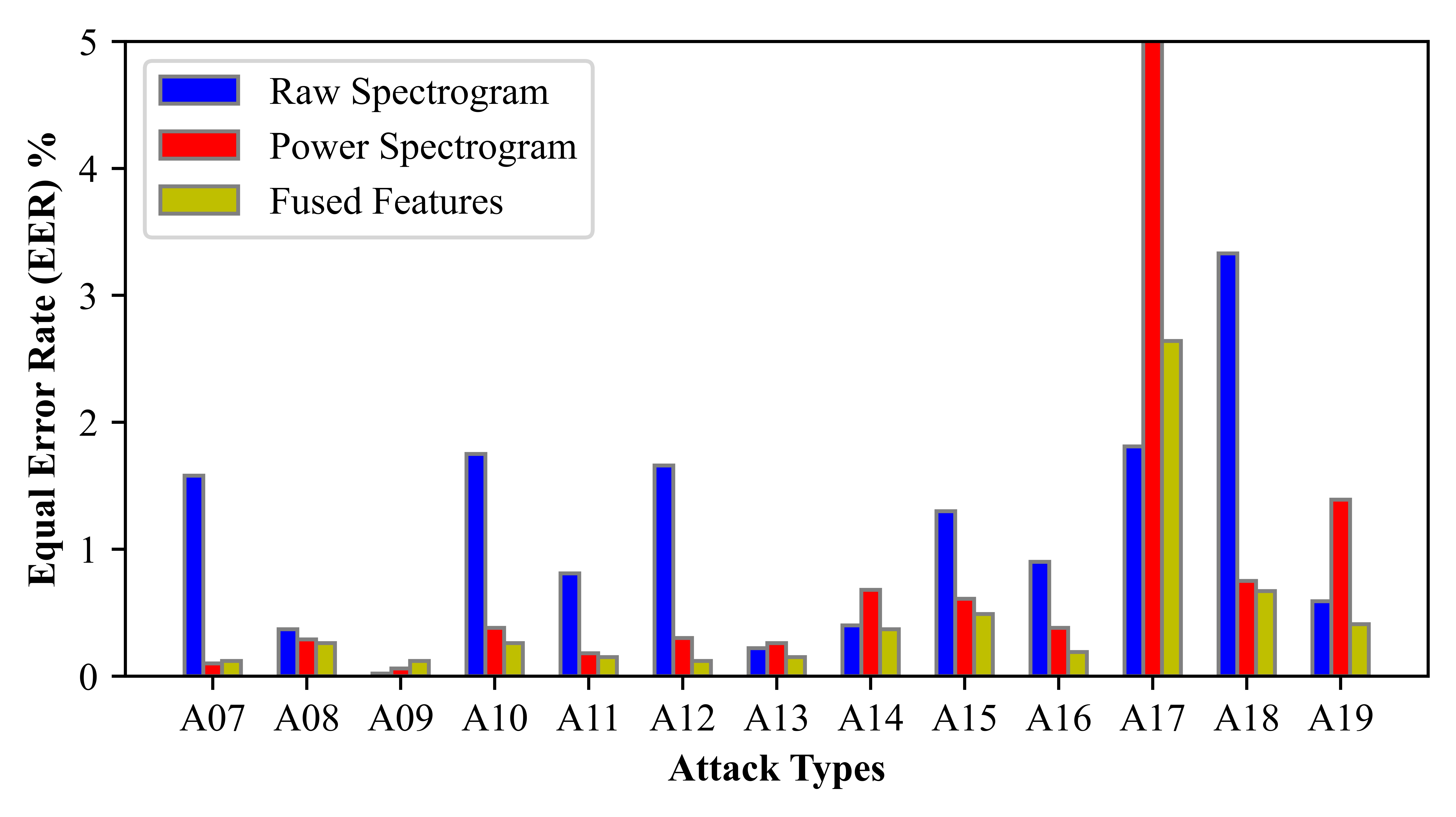}
  \caption{Illustration of the performance for anti-spoofing on ASVspoof2019 LA Challenge, which is highly sensitive with the order of the spectral features used. }
  \label{fig:lfcc_raw}
\end{figure}

\begin{figure*}[h]
  \centering
  \includegraphics[width=0.95\linewidth]{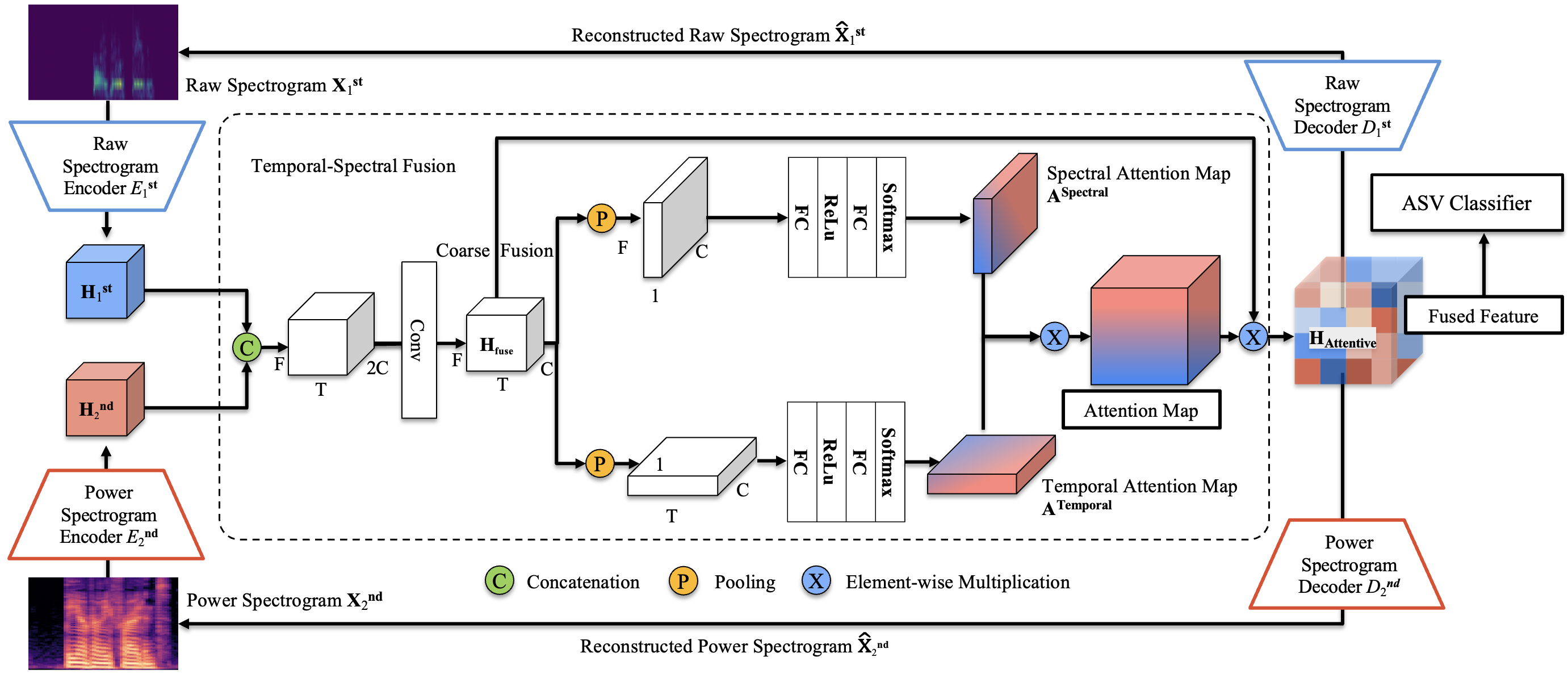}
  \caption{Illustration of the overall architecture for the proposed $\text{S}^2\text{pecNet}$.}
  \label{fig:overall_model}
\end{figure*}

These existing methods are often based on a specific category of audio features. 
However, as shown in Figure~\ref{fig:lfcc_raw}, different types of features exhibit varying effectiveness in detecting different types of attacks. Thus, instead of using a single source of spectral features as in AASIST, we suggest that diverse orders of audio spectral patterns can benefit the speech anti-spoofing in a complementary manner. For example, the $2^\text{nd}$-order spectrograms (i.e., power spectrograms) are suggested to be more sensitive to the noise patterns in real-world speech \cite{tyagi2005desensitizing}.
As shown in Figure 3, the power spectrogram can detect subtle variations regarding spoofing clues in high frequency regions with low amplitude values, compared with the raw spectrogram.

\begin{figure}[h]
  \centering
  \includegraphics[width=\linewidth]{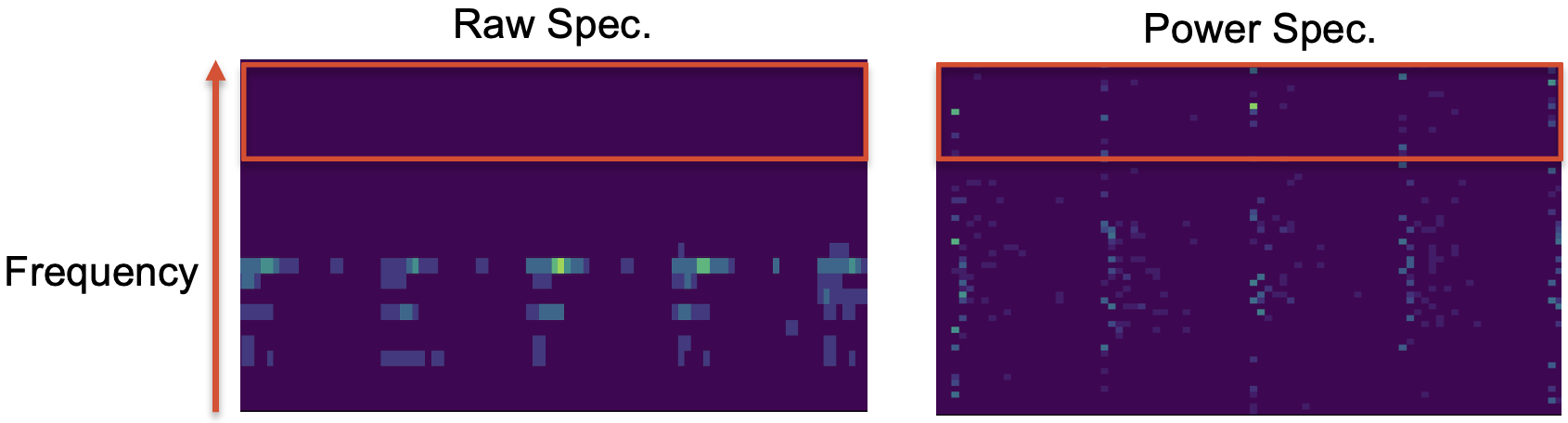}
  \caption{Illustration of raw and power spectrograms, where the area in red bounding boxes represents high frequency regions.}
  \label{fig:r-p}
\end{figure}

Therefore, in this study, a novel deep learning architecture, namely $\text{S}^2\text{pecNet}$, is proposed for robust anti-spoofing by using multi-order spectrogram patterns, which are up to the second-order including both raw and power spectrograms. 
Specifically, raw and power spectral patterns are fused in a coarse-to-fine manner and two branches are involved for the fine-level fusion from the spectral and temporal contexts. 
To minimize the information loss during the feature learning and fusion, a reconstruction mechanism is devised to reconstruct the fused representation to its associated input spectrograms.
Comprehensive experiments on a commonly used dataset - ASVspoof 2019 LA,  demonstrates the effectiveness of our proposed method, $\text{S}^2\text{pecNet}$, which achieves the state-of-the-art performance regarding the metrics minimum tandem detection cost function (min t-DCF) \cite{kinnunen2020tandem} and equal error rate (EER) \cite{1409642}.

In summary, the key contributions of this work are: (i) a novel deep learning based fusion architecture for audio anti-spoofing with multi-order spectrograms; (ii)
a coarse-to-fine fusion mechanism with two branches that are involved for the fine-level fusion from the spectral and temporal contexts; and 
(iii) a reconstruction strategy to maintain the information in the fused speech representations. 

\section{Proposed Method}

Figure~\ref{fig:overall_model} illustrates the overall architecture of the $\text{S}^2\text{pecNet}$ method.
The $1^\text{st}$-order raw spectrogram and the $2^\text{nd}$-order power spectrogram of an input audio are first fed into their encoders, respectively.
Next, the two encoded features are concatenated and fed into a temporal-spectral fusion module in pursuit of a spoofing-sensitive representation by exploiting the supplementary information between the two spectrograms. 
Additionally, to minimise the potential information loss of the fused representations, a reconstruction mechanism is introduced with two decoders to reconstruct the fused features back to the original raw and power spectrograms. 

\subsection{Raw Spectrogram and Power Spectrogram Encoding}
$\text{S}^2\text{pecNet}$ takes an input audio waveform $\mathbf{X}$ from its first-order and second-order spectral characteristics as input in pursuit of a comprehensive representation of various spoofing patterns. 
For the first-order patterns, the input audio's raw spectrogram $\mathbf{X}_{1^\textbf{st}}$ is fed into a CNN based encoder $E_{1^\textbf{st}}$ to formulate an audio feature map $\mathbf{H}_{1^\textbf{st}}\in \mathbb{R}^{C \times F \times T}$, where $C$, $F$, and $T$ denote the number of channels, the number of spectral bins, and sequence length, respectively. 
In terms of the second-order patterns, the input audio's power spectrogram $\mathbf{X}_{2^\textbf{nd}}$ is encoded by another CNN encoder $E_{2^\textbf{nd}}$ and a feature map can be obtained as $\mathbf{H}_{2^\textbf{nd}}\in \mathbb{R}^{C \times F \times T}$. Note that $E_{1^\textbf{st}}$ and $E_{2^\textbf{nd}}$ are set to generate their output feature maps with the same dimension. 

\subsection{Temporal-Spectral Fusion}

The output feature maps $\mathbf{H}_{1^\textbf{st}}$ and $\mathbf{H}_{2^\textbf{nd}}$ of the two encoders are with different spectral orders. Therefore, a temporal-spectral fusion (TSF) module is devised to refine and fuse the two feature maps. TSF formulates the dependencies between the two spectral domains and explores their complementary spoofing-related patterns in a coarse-to-fine manner. 
Initially, a coarse fusion step is performed by concatenating $\mathbf{H}_{1^\textbf{st}}$ and $\mathbf{H}_{2^\textbf{nd}}$ in a channel-wise manner and applying a set of convolution filters on the concatenated feature map to obtain a coarse fused representation $\mathbf{H}_{\text{fuse}}$. Then, to characterise finer spoofing-sensitive features from $\mathbf{H}_{\text{fuse}}$, an attention map $\mathbf{A}$ is obtained to highlight its patterns that are more susceptible to spoofing by formulating the long-term temporal dependencies and the spectral patterns. 

To obtain $\mathbf{A}$, two sub-attention maps $\mathbf{A}^\text{spectral}$ and $\mathbf{A}^{\text{temporal}}$ are derived from two different contexts: one explores the temporal context while the other explores the spectral context. In detail, $\mathbf{H}_{\text{fuse}}$ is pooled along its temporal and spectral dimensions, respectively, and we have:
\begin{equation}
  \mathbf{H}_\text{fuse}^\text{spectral} = \text{max}_t(|\mathbf{H}_{\text{fused}}|),
  \mathbf{H}_\text{fuse}^\text{temporal} = \text{max}_s(|\mathbf{H}_{\text{fused}}|),
  \label{eq2}
\end{equation}
where $\mathbf{H}_\text{fuse}^\text{spectral} \in \mathbb{R}^{C \times F \times 1}$, $\mathbf{H}_\text{fuse}^\text{temporal} \in \mathbb{R}^{C \times 1 \times T}$,  $|\cdot|$ refers to an element-wise absolute operator, $\text{max}_s$ is a global spectral pooling operator and $\text{max}_t$ indicates a global temporal pooling operator. To this end, $\mathbf{H}_\text{fuse}^\text{spectral}$ contains global temporal information across frequency bins, and $\mathbf{H}_\text{fuse}^\text{temporal} $ contains global spectral information across time.
Next, the two attention maps $\mathbf{A}^\text{spectral} \in \mathbb{R}^{C \times F \times 1}$ and $\mathbf{A}^\text{temporal} \in \mathbb{R}^{C \times 1 \times T}$ are obtained as:
\begin{equation}
  \mathbf{A}^{\text{spectral}} = \text{Conv}_{\text{s}}(\mathbf{H}_\text{fuse}^\text{spectral}) ,
  \mathbf{A}^{\text{temporal}} = \text{Conv}_{\text{t}}(\mathbf{H}_\text{fuse}^\text{temporal} ),
  \label{eq4}
\end{equation}
where $\text{Conv}_{\text{t}}$ and $\text{Conv}_{\text{s}}$ denote the convolution layers for obtaining the two attention maps, respectively.
To this end, the final attention map $\mathbf{A}$ is obtained by: $\mathbf{A} = \mathbf{A}^\text{spectral}\times\mathbf{A}^\text{temporal} $, and the final fused representation can be derived as:
\begin{equation}
  \mathbf{H}_\text{attentive} = \mathbf{A}\times \mathbf{H}_{\text{fused}}.
  \label{eq5}
\end{equation}

\subsection{Raw Spectrogram and Power Spectrogram Decoding}

To prevent information loss during the encoding procedure and the feature fusion, a raw spectrogram decoder $D_{1^\textbf{st}}$ and a power spectrogram decoder $D_{2^\textbf{nd}}$ are devised to reconstruct the input raw spectrograms and power spectrograms, respectively. Specifically, $D_{1^\textbf{st}}$ consists of  a series of deconvolution layers, which takes the fused feature $\mathbf{H}_\text{attentive}$ to reconstruct the raw spectrogram $\hat{\mathbf{X}}_{1^\textbf{st}}$. Similarly, $D_{2^\textbf{nd}}$ applies deconvolution layers to $\mathbf{H}_\text{attentive}$ and produce the reconstructed power spectrogram $\hat{\mathbf{X}}_{2^\textbf{nd}}$. A raw spectrogram reconstruction loss $\mathcal{L}_{1^\textbf{st}}$ and a power spectrogram reconstruction loss $\mathcal{L}_{2^\textbf{nd}}$ are introduced as:
\begin{equation}
  \mathcal{L}_{1^\textbf{st}} = \Vert \hat{\mathbf{X}}_{1^\textbf{st}} - \mathbf{X}_{1^\textbf{st}}\Vert, 
  \mathcal{L}_{2^\textbf{nd}} = \Vert \hat{\mathbf{X}}_{2^\textbf{nd}} - \mathbf{X}_{2^\textbf{nd}} \Vert.
  \label{eq6}
\end{equation}
Minimizing the two losses aims to best reconstruct the original raw and power spectrograms using the fused representation.

\subsection{Spoofing Detection}
A classifier further takes the fused representation $\mathbf{H}_\text{attentive}$ as input to produce a binary classification prediction $\hat{y}$. The classification loss $\mathcal{L}_\text{cls}$ is with a weighted binary cross-entropy (WACE) to quantify the difference between the prediction $\hat{y}$ and the ground truth $y$, which can be formulated as:
\begin{equation}
\mathcal{L}_\text{cls}=-\frac{1}{N} \sum_{i=1}^N y \cdot \log (\hat{y})+(1-y) \cdot \log (1-\hat{y}).
\label{eq8}
\end{equation}

\subsection{Model Training}
The overall loss $\mathcal{L}$ of the proposed method $\text{S}^2\text{pecNet}$ is with the three loss terms mentioned above, including the raw spectrogram reconstruction loss $\mathcal{L}_{1^\textbf{st}}$, the power spectrogram reconstruction loss $\mathcal{L}_{2^\textbf{nd}}$, and the classification loss $\mathcal{L}_\text{cls}$. The relative importance of them is controlled by a hyper-parameter $\alpha$:
\begin{equation}
  \mathcal{L} = \alpha (\mathcal{L}_{1^\textbf{st}} + \mathcal{L}_{2^\textbf{nd}}) + \mathcal{L}_\text{cls}.
  \label{eq9}
\end{equation}

\section{Experiments \& Discussions}

\subsection{Dataset and Evaluation Metrics}
A widely used dataset, ASVspoof2019 LA \cite{todisco2019asvspoof}, was adopted for evaluation. It consists of both bona fide audio recordings and 19 different types of spoofing attacks generated through text-to-speech (TTS) \cite{shchemelinin2013examining} and voice conversion (VC) \cite{kinnunen2012vulnerability}. We followed the same partitions of training, development, and evaluation as in AASIST. The training and development partitions include 6 different spoofing attacks (A01-A06), while the evaluation partition includes 13 different attacks (A07-A19). The overview of this dataset is listed in Table~\ref{tab:la_challenge}. For evaluation metrics, the minimum tandem detection cost function (min t-DCF) and the equal error rate (EER) were adopted. 
Moreover, it has been demonstrated that the performance of spoofing detection algorithms can vary greatly depending on initial random seeds~\cite{wang2021comparative}. Hence, we report the average metrics across a number of random seeds.

\begin{table}[H]
\caption{Overview of the ASVspoof2019 LA dataset}
\label{tab:la_challenge}
\centering
\begin{tabular}{l|c|c|c} 
\hline
Partition& Bona fide & \multicolumn{2}{c}{ Spoofed } \\
\cline { 2 - 4 } & \# utterance & \# utterance & attacks \\
\hline Training & 2,580 & 22,800 & A01 - A06 \\
Development & 2,548 & 22,296 & A01 - A06 \\
Evaluation & 7,355 & 63,882 & A07 - A19\\
\hline
\end{tabular}
\end{table}

\subsection{Implementation Details}

A raw waveform was obtained with 64,600 frames (approximately 4 seconds). A sinc-convolution filter was used to obtain the raw spectrogram of an input audio, which was further encoded through a ResNet encoder with 6 residual blocks. For the power spectrogram, we formulated a 60-dimensional LFCC features for a frame, of which the size is 20 ms with a hop size of 10 ms.
A ResNet-18 was utilised as the power spectrogram encoder. The classifier for spoofing detection followed the setting as in AASIST. Both $\text{CONV}_{\text{s}}$ and $\text{CONV}_{\text{t}}$ of the TSF module consisted of two fully-connected layers, a batch normalization layer, a SiLU function and a sigmoid function. Our $\text{S}^2\text{pecNet}$ was implemented and trained using PyTorch for 100 epochs on an NVIDIA RTX A6000 GPU with a batch size 48. An Adam optimizer was adopted with a learning rate of $3\times10^{-4}$ and a cosine annealing learning rate decay.

\begin{table}[H]
\caption{Comparison with the state-of-the-art methods. min t-DCF and EER values were from AASIST, while \# of parameters and inference time were based on official implementations, indicating the average inference duration for one-second audio.}
\label{tab:sota_comparison}
\setlength{\tabcolsep}{3pt}
\centering
\begin{tabular}{lccccc}
\hline Method & \#Param  & Runtime & min t-DCF $\downarrow$ & EER $\downarrow$ \\
\hline  
AASIST & 297k  &0.0052 & $0.028$ & $0.83$ \\
RawGAT-ST \cite{tak2021end}& 437k  &0.0049 &$0.034$ & $1.06$ \\
MCG-Res2Net \cite{li2021channel}& 960k  & - & $0.052$ & $1.78$ \\
OC-Softmax \cite{zhang2021one}& 12450k  & 0.0018 & $0.059$ & $2.19$ \\
SE-Res2Net \cite{li2021replay}& 920k  & - & $0.074$ & $2.50$ \\
\hline
$\text{S}^\text{2}\text{pecNet}$ (Ours) & 1284k  &  0.0072 &\textbf{0.024} & \textbf{0.77} \\
\hline
\end{tabular}
\end{table}

\subsection{Performance Comparison}

Table~\ref{tab:sota_comparison} lists a performance comparison between our proposed $\text{S}^2\text{pecNet}$ and the state-of-the-art methods. It can be observed that $\text{S}^2\text{pecNet}$ outperforms these existing methods and shows strong capability for robust audio spoofing detection.
Specifically, Table~\ref{tab:aasist_comparison} 
lists the comparison between our $\text{S}^2\text{pecNet}$ and the state-of-the-art AASIST method regarding the best metrics. Our $\text{S}^2\text{pecNet}$ demonstrates superior performance in terms of both min t-DCF and EER, where our method improves the average EER by 25\%. In addition, $\text{S}^2\text{pecNet}$ has superior or comparable performance on most attacks, except for the A17 attack where the AASIST model outperforms ours with a significant gap. 
A17 employed an acoustic model VAE-GAN~\cite{hsu2017voice}, which generate spectral shapes that are more realistic and detailed in the high-frequency patterns. The power spectrogram can be fooled by A17 since its modelling relies on observing subtle variations in high frequency regions.

\begin{table*}[th]
\caption{Comparison with AASIST. Results are based on EER and values in parentheses show the best readout. }
\label{tab:aasist_comparison}
\centering
\setlength{\tabcolsep}{4pt}
\begin{tabular}{l|ccccccccccccc|cc}
\hline System & A07 & A08 & A09 & A10 & A11 &A12& A13 & A14 & A15 & A16 & A17 & A18 & A19 & min t-DCF & EER (\%)  \\
\hline AASIST & 0.80 & 0.44 & $\textbf{0.00}$ & 1.06 & 0.31 & 0.91 & $\textbf{0.10}$ & $\textbf{0.14}$ & 0.65& 0.72 & $\textbf{1.52}$ & 3.40 & 0.62 &0.035(0.028) & $1.13(0.83)$  \\
\hline Ours & $\textbf{0.12}$ & $\textbf{0.26}$ & $0.12$ & $\textbf{0.26}$ & $\textbf{0.15}$ & $\textbf{0.12}$ & $0.15$ & 0.37 & $\textbf{0.49}$ & $\textbf{0.19}$ & 2.64 & $\textbf{0.67}$ & $\textbf{0.41}$ &$\textbf{ 0.027(0.024)}$ & $\textbf{0.84(0.77)}$  \\
\hline
\end{tabular}
\end{table*}

\subsection{Ablation Study}
\subsubsection{Spectral Complementary Patterns}

An ablation study was conducted to investigate the impact of the complementary information between the two spectrograms. Three settings were investigated under the same condition with raw spectrograms, power spectrograms and fused spectrograms (i.e., the representations obtained from S$^{2}$pecNet), respectively. 

\begin{figure}[h]
  \centering
  \includegraphics[width=0.9\linewidth]{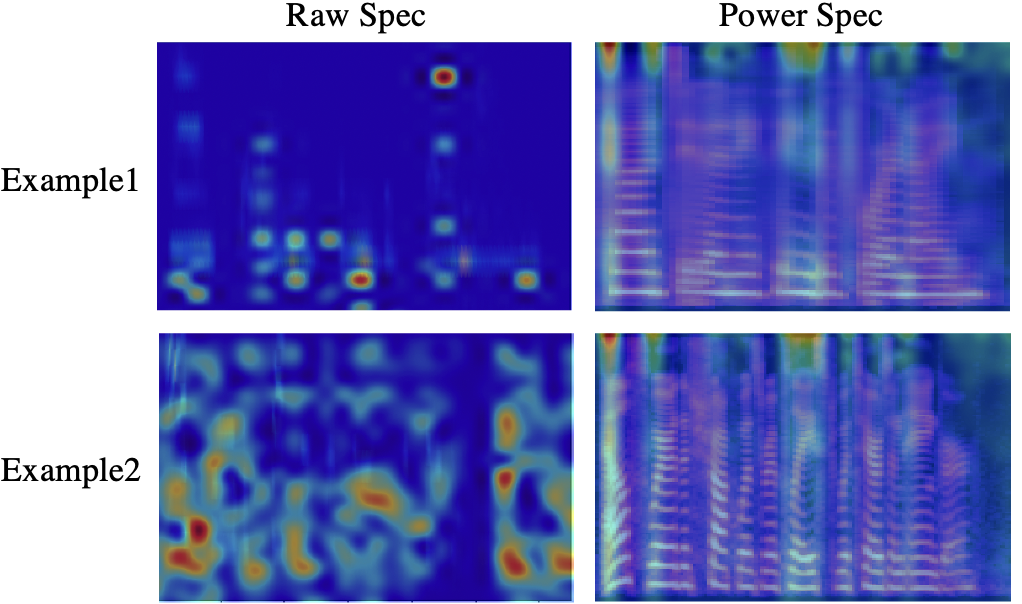}
  \caption{Grad-CAM on spectrograms of two examples.}
  \label{fig:cam}
\end{figure}

As listed in Table~\ref{tab:individual_attack_comparision}, raw spectrogram is able to achieve significantly better performance than power spectrogram on A17 and A19, while the power spectrogram demonstrates largely better performance on A07, A10, A11, A12, A15, A16, and A18. Since each setting demonstrated its advantages on different attacks, the fused spectrograms can effectively exploit the complementary information of them to achieve the best performance in most attacks.
Grad-CAM~\cite{selvaraju2017grad} was adopted to visualise the attention maps on the two spectrograms. 
As shown in Figure~\ref{fig:cam}, $\text{S}^2\text{pecNet}$ is able to identify high frequency information from the power spectrogram, while focusing on low frequency information from the raw spectrogram, which confirms that utilising two spectrograms can explore the complementary information together for better spoofing detection.

\begin{table}[th]
\caption{Performance comparisons regarding EER on raw spectrograms, power spectrograms, and fused spectrograms.}
\label{tab:individual_attack_comparision}
\centering
\begin{tabular}{c|ccc}
 \hline Attacks $\downarrow$ & Raw spec.
& Power spec. & Fused spec.\\
\hline A07 & $1.58$ & $\textbf{0.10}$ & $0.12$ \\
A08 & $0.37$ & $0.29$ & $\textbf{0.26}$ \\
A09 & $\textbf{0.02}$ & $0.06$ & $0.12$ \\
A10 & $1.75$ & $0.48$ & $\textbf{0.26}$ \\
A11 & $0.81$ & $0.18$ & $\textbf{0.15}$ \\
A12 & $1.66$ & $0.30$ & $\textbf{0.12}$ \\
A13 & $0.22$ & $0.26$ & $\textbf{0.15}$ \\
A14 & $0.40$ & $0.68$ & $\textbf{0.37}$ \\
A15 & $1.30$ & $0.61$ & $\textbf{0.49}$ \\
A16 & $0.90$ & $0.38$ & $\textbf{0.19}$ \\
A17 & $\textbf{1.81}$ & $31.46$ & $2.64$ \\
A18 & $3.33$ & $0.75$ & $\textbf{0.67}$ \\
A19 & $0.59$ & $1.39$ & $\textbf{0.41}$ \\
\hline
\end{tabular}
\end{table}

\subsubsection{Impact of TSF Module}

To evaluate the effectiveness of the TSF module, comparisons were conducted with other fusion methods, including concatenation, early fusion, late fusion, and other state-of-the-art fusion methods. As shown in Table~\ref{tab:fusion_comparison}, concatenation can achieve reasonably good performance by avoiding information loss, whereas many complex fusion methods lead to worse performance, which suggest greater information loss. Our TSF module can effectively exploit the complementary information between two spectral domains embedded with temporal dependencies, and achieved the best performance.

\begin{table}[h]
\caption{Comparisons with different fusion methods.}
\label{tab:fusion_comparison}
\centering
\begin{tabular}{lcc}
\hline Method & EER(\%) $\downarrow$ & $\min$ t-DCF  $\downarrow$\\
\hline 
Early-Fusion \cite{liu2018efficient} & $7.90$ & $0.1800$ \\
Late-Fusion \cite{zhang2019late}& $3.53$ & $0.0848$ \\
MFFN \cite{zheng2023mffn} & $3.24$ & $0.0645$ \\
CEFNet(ACM) \cite{feng2021encoder} & $1.48$ & $0.0465$ \\
SA-Fuser(w) \cite{zhong2023anticipative} & $1.32$ & $0.0320$ \\
Concatenation & $1.03$ & $0.0340$ \\
\hline
TSF (Ours) & $\textbf{0.84}$ & $\textbf{0.0271}$ \\

\hline
\end{tabular}
\end{table}

\subsubsection{Impact of Reconstruction Decoders}
 
We further investigate the impact of the reconstruction decoders which aim to retain the original information in the final fused representation. As shown in Table~\ref{tab:decoder_comparison}, the reconstruction decoders can improve detection performance by retaining more useful complementary information.
Additionally, the settings of hyper-parameter $\alpha$ are explored, which is used to balance the detection and the reconstruction. The results in Table~\ref{tab:alpha_beta} indicate that the best performance is achieved with $\alpha = 0.1$.

\begin{table}[h]
\caption{Ablation study on reconstruction decoders.}
\label{tab:decoder_comparison}
\centering
\begin{tabular}{lcc}
\hline Method & EER(\%) $\downarrow$ & $\min$ t-DCF  $\downarrow$\\
\hline w/o reconstruction decoders & $0.93$ & $0.0295$ \\
w/ reconstruction decoders & $\textbf{0.84}$ & $\textbf{0.0271}$ \\
\hline
\end{tabular}
\end{table}

\begin{table}[h]
\caption{Hyper-parameter selection in terms of $\alpha$.}
\label{tab:alpha_beta}
\centering
\begin{tabular}{lcc}
\hline $\alpha$ & EER(\%) $\downarrow$ & $\min$ t-DCF  $\downarrow$\\
\hline 1 & $1.02$ & $0.0295$ \\
$0.1$ & $\textbf{0.84}$ & $\textbf{0.0271}$ \\
0.01 & $0.96$ & $0.0283$ \\
\hline
\end{tabular}
\end{table}

\section{Conclusion}

We present a novel method $\text{S}^2\text{pecNet}$ for audio spoofing detection by exploiting complementary information from multi-order spectrograms. Specifically, a TSF module is devised to fuse the two spectral representations in a coarse-to-fine manner. To minimize information loss, a fused representation is reconstructed to its input spectrograms. Comprehensive experiments demonstrate the superiority of $\text{S}^2\text{pecNet}$ over the state-of-the-art methods. As $\text{S}^2\text{pecNet}$ does not work well for some specific spoofing attacks, utilising higher-order and flexible spectral patterns in a data-driven scheme could be worth studying in future research. Additionally, considerations should be given to advanced techniques for spurious synthesis (e.g., \cite{liu2023making}), to enhance the robustness of audio spoofing detection. 

\newpage
\bibliographystyle{IEEEtran}
\bibliography{mybib}
\end{document}